\documentclass[aps,prb] {revtex4}
\usepackage{graphicx}
\usepackage{amsfonts,amsmath}
\begin{document}
\title{Thermopower Oscillation Symmetries in a \\Double-Loop Andreev Interferometer}

\author{P. Cadden-Zimansky, Z. Jiang and V. Chandrasekhar}

\affiliation{Department of Physics and Astronomy, Northwestern University, 2145 Sheridan Road, Evanston, IL 60208, USA} 

\date{\today}

\begin{abstract}
Andreev interferometers, normal metal wires coupled to superconducting loops, display phase coherent changes as the magnetic flux through the superconducting loops is altered.  Properties such as the electronic and thermal conductance of these devices have been shown to oscillate symmetrically about zero with a period equal to one superconducting flux quantum, $\Phi_o = h/2e$.  However, the thermopower of these devices can oscillate symmetrically or antisymmetrically depending on the geometry of the sample, a phenomenon not well understood theoretically.  Here we report on thermopower measurements of a double-loop Andreev interferometer where two Josephson currents in the normal metal wire may be controlled independently.  The amplitude and symmetries of the observed thermopower oscillations may help to illuminate the unexplained dependence of oscillation symmetry on sample geometry.



\end{abstract}

\maketitle


\section{Introduction and Background}
While studies of normal metal-superconductor (NS) heterostructures in the proximity regime have seen a great progression of understanding over the past few decades due to improved fabrication and experimental techniques and the success of the quasi-classical Usadel formulation, there remain several quantitative and qualitative mysteries that have eluded explanation.  One of these is the observed thermoelectric effects of a NS device known as an Andreev interferometer, which consists of a superconducting loop which is completed by a section of a normal metal wire short enough to be able to support a Josephson current.  When a supercurrent is created in the section of the normal wire by threading a flux through the superconducting ring, the electrical and thermal transport properties of the normal wire have been shown to fluctuate with a period of one superconducting flux quantum, $\Phi_0 = h/2e$.  As the electrical and thermal conductance of the normal wire should be dependent, by arguments of symmetry, only on the magnitude of the Josephson current and not on its direction, the oscillations of these quantities as a function of flux through the superconducting ring will be symmetric, a prediction confirmed by experiment \cite{pothier} \cite{jiang}.  However, surprisingly, it has been observed that the oscillations of the thermopower of the wire can be either symmetric or antisymmetric with flux depending upon the geometry of the sample.

In the linear approximation of small voltage, $\Delta{V}$, and temperature difference, $\Delta{T}$, applied to a wire, the current through the wire is given by \cite{ashcroft}:
\begin{equation}
\label{eqn1}
I = G\Delta{V} + \eta\Delta{T}
\end{equation}
In the usual case of an applied voltage the transport is dominated by the electrical conductance, $G$, but one may observe the effect of the thermoelectric term $\eta$ under the conditions that there is a temperature gradient and no electrical current is allowed to flow.  Under these conditions the thermal gradient will induce a voltage across the wire and one can measure the thermopower, $S\equiv\Delta{V}/\Delta{T}=-{\eta}/{G}$.  It should be noted that since $G$ oscillates symmetrically with flux for Andreev interferometers, the symmetry of the thermopower with flux for these devices will be determined by the thermoelectric coefficient $\eta$.

\begin{figure}
\begin{center}
\includegraphics[width=11cm]{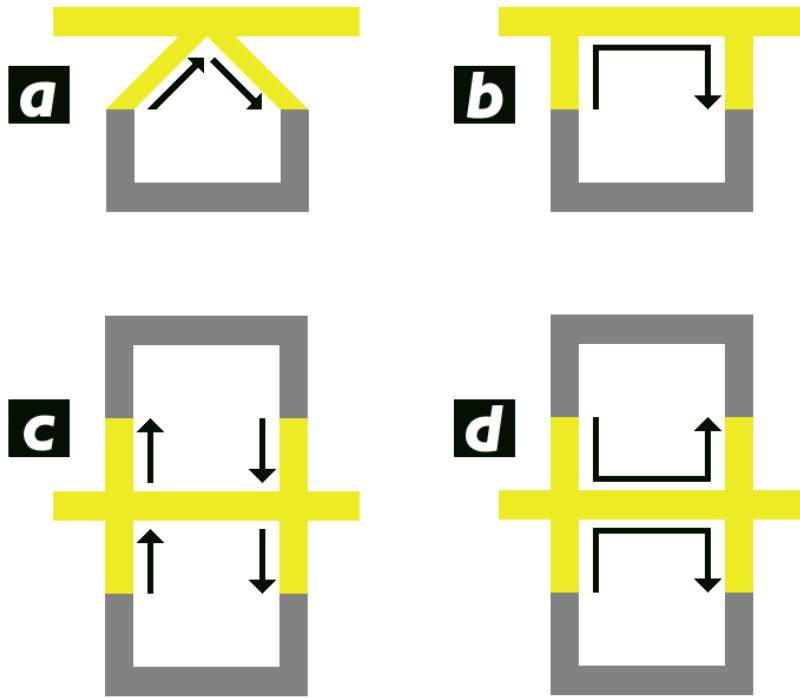}
\caption{Schematics of sample geometries previously investigated for thermopower symmetry dependence.  In all geometries a horizontal normal wire (Au) is used to complete one or two superconducting (Al) loops.  A temperature gradient is established along the normal wire and a Josephson current may be sent along or across a section of the wire by threading a magnetic flux through the superconducting loop(s) while the thermopower of the wire is measured.  (a) The `house' geometry, where the Josephson current intersects the thermal gradient along the normal wire at a single point, displays symmetric oscillations in the thermopower.  (b)  The `parallelogram' geometry, where the Josephson current travels with the thermal gradient along a section of the normal wire, displays antisymmetric oscillations.  (c)  The `in-phase' current configuration for an in-line, double-loop interferometer displays symmetric oscillations in the thermopower.  (d)  The `out-of-phase' current configuration for the same interferometer displays antisymmetric oscillations.}
\label{fig1}
\end{center}
\end{figure}

Measurements of the thermopower of Andreev interferometers initially focused on two types of geometries shown in Fig. \ref{fig1}(a) and (b) known as the `house' and `parallelogram' geometry respectively \cite{eom}.  In both cases a thermal gradient is established along the normal metal wire and the induced voltage across the wire is measured as magnetic flux is threaded through the superconducting ring.  For the house geometry the Josephson current intersects the thermal current traveling along the normal wire at a single point and symmetric thermopower oscillations are observed.  For the parallelogram geometry the Josephson current flows with the thermal current along a section of the normal wire and antisymmetric thermopower oscillations are observed.  Theoretical investigations using the Usadel formulation of the BCS model of superconductivity have predicted the correct thermopower symmetry and reasonable thermopower magnitudes for the parallelogram geometry \cite{virtanen, zou}, while an understanding of the symmetric oscillations of the house geometry remains elusive.  

\begin{figure}
\begin{center}
\includegraphics[width=16cm]{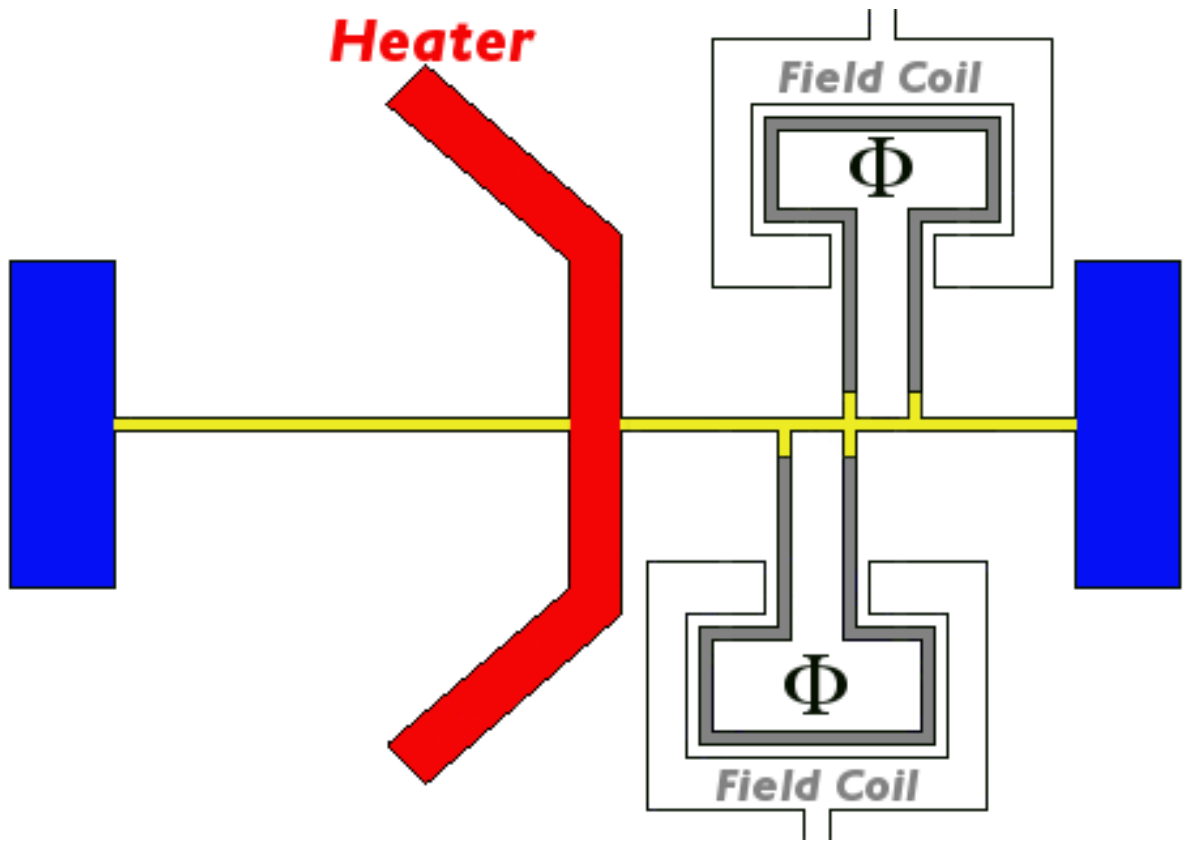}
\caption{Schematic of the offset double-loop sample measured.  Current is sent through a gold heater line running vertically through the center of the sample.  This wire creates a thermal gradient from the center of the sample across the horizontal, normal (Au) wires at left and right to the two cold reservoirs at the far ends.  The horizontal wire on the left is a control wire.  The wire on the right is attached to two offset superconducting (Al) loops to form an Andreev interferometer.  Supercurrents are established in these two loops by threading a local magnetic flux, $\Phi$, through them.  The flux through each loop may be controlled independently using two on-chip field coils.  Typically the magnitude of the flux through each loop is made identical, though the polarity may be reversed.  If the established supercurrents circulate in the same direction (e.g. both clockwise) the measurement is said to be `in-phase', if they circulate in the opposite direction they are `out-of-phase'.  The symmetry and relative amplitude of the thermopower oscillations as a function of flux through the loops are measured using the second derivative technique of Ref. \cite{dikin}.}
\label{fig2}
\end{center}
\end{figure}

In order to illuminate the roles that the supercurrent path and thermal current path, along with associated quasiparticle current and branch imbalance, play in determining the symmetry of the thermopower oscillations, our group previously fabricated and measured a double-loop interferometer with each superconducting loop being completed along the same section of normal metal wire (Fig. \ref{fig1}(c) \& (d)) \cite{jiang2}.  As the flux through each loop could be controlled independently, two different supercurrent configurations could be realized in the same device.  Fig. \ref{fig1}(d) shows the `out-of-phase' configuration where the supercurrents from each loop travel in the same direction along the thermal gradient.  This configuration leads to asymmetric thermopower oscillations, which is unsurprising since it is essentially a doubling of the simple parallelogram geometry.  The `in-phase' configuration of Fig. \ref{fig1}(c) leads to symmetric thermopower oscillations.  However it is not clear whether this configuration should be interpreted as involving two supercurrents traveling in opposite directions along the thermal gradient of the normal wire or whether the supercurrents should be seen as bypassing the central normal wire section and only intersecting the normal wire at two points, effectively a doubling of the house geometry.  To further clarify the supercurrent path we have fabricated and measured a variation of this double-loop geometry shown in Fig. \ref{fig2}.  In this device the loops are offset rather than in-line so that the in-phase configuration results in supercurrents in opposite directions in the normal wire, but avoids the potential for supercurrent not flowing at all along the thermal gradient, as was possible in the previous device, by sending the opposing supercurrents through adjacent sections of the normal wire.

\section{Sample Fabrication and Measurement}

\begin{figure}
\begin{center}
\includegraphics[width=14cm]{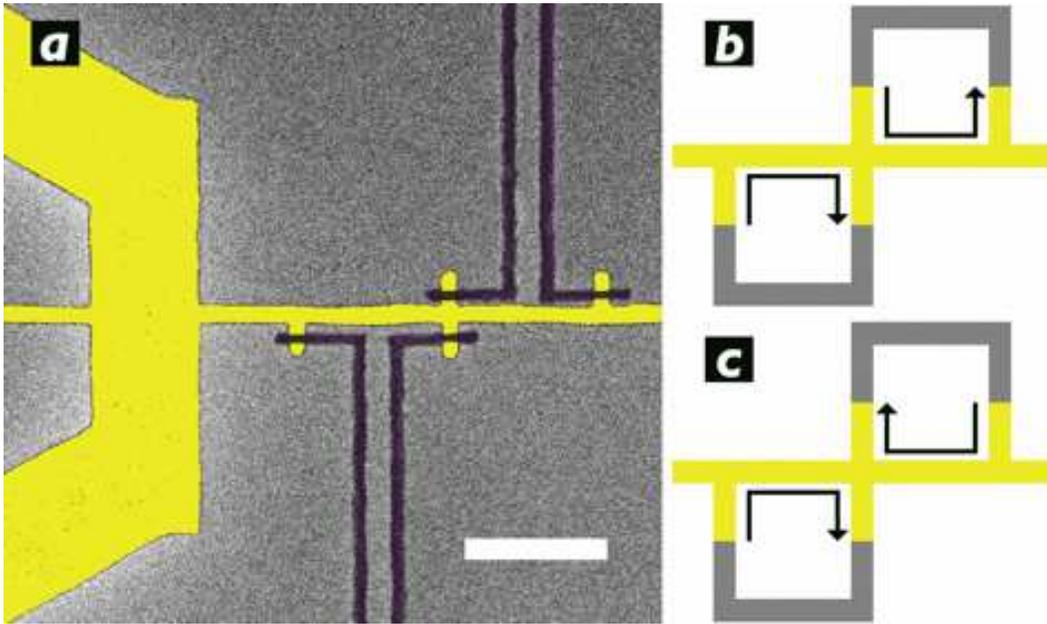}
\caption{(a) Colorized scanning electron micrograph of the section of the sample where the two superconducting loops are completed along the normal metal wire.  On the left side is the wide heater line used to establish a thermal gradient.  The size bar is 1 $\mu$m.  (b)  Schematic of the supercurrent path along the normal wire for the `out-of-phase' configuration.  (c)  Schematic of the `in-phase' configuration.}
\label{fig3}
\end{center}
\end{figure}

The device, shown schematically in Fig. \ref{fig2}, with a scanning electron micrograph of the NS intersections shown in Fig. \ref{fig3}(a), was fabricated by multi-level e-beam lithography on a SiO$_2$/Si substrate.  The normal metal (50 nm of 99.999\% pure Au) was deposited first using an e-gun evaporator followed by the superconducting layer (80 nm of 99.999\% pure Al).  An \textit{in situ} Ar$^+$ plasma etch was performed prior to the second deposition to ensure clean contacts between the two layers.  Four-terminal measurements of the sample were made in an Oxford dilution refrigerator using a lock-in amplifier and modified Adler-Jackson resistance bridge.  At 20 mK the resistivity of the Au was measured to be $\rho_{Au}=2.1\;\mu\Omega\cdot$cm.  Using the textbook values \cite{ashcroft} for the Fermi velocity and electronic mean free path-resistivity product for Au, $v_f=1.4\times10^8$ cm/sec and ${l_e}\,\rho_{Au}=8.3\;\mu\Omega\cdot${cm}$^2$, we calculate an Au diffusion constant of $D = 187$ cm$^2$/sec for our sample.

A wide Au heater line controlled by Joule heating that runs through the center of the sample establishes a temperature gradient with two cool reservoirs at either end which are maintained at the mixing chamber temperature.  The gradient runs along the offset double-loop Andreev interferometer on the right side and along a plain normal wire of equal length on the left.  The flux through the two interferometer loops may be controlled independently by sending a DC current through two on-chip field coils.  While resistance measurements of the sample are made directly, the thermopower measurements are made using the second derivative technique described in Ref. \cite{dikin}.  A 2 $\mu$V, 42 Hz signal is sent along the heater line and the voltage between the two cool reservoirs is measured using a lock-in amplifier set at 84 Hz.  This voltage does not give us a quantitative measurement of the thermopower, since the temperature is not directly measured, but it does provide information about the thermopower's symmetry and relative amplitude.
 
\section{Experimental Results}

\begin{figure}
\begin{center}
\includegraphics[width=16cm]{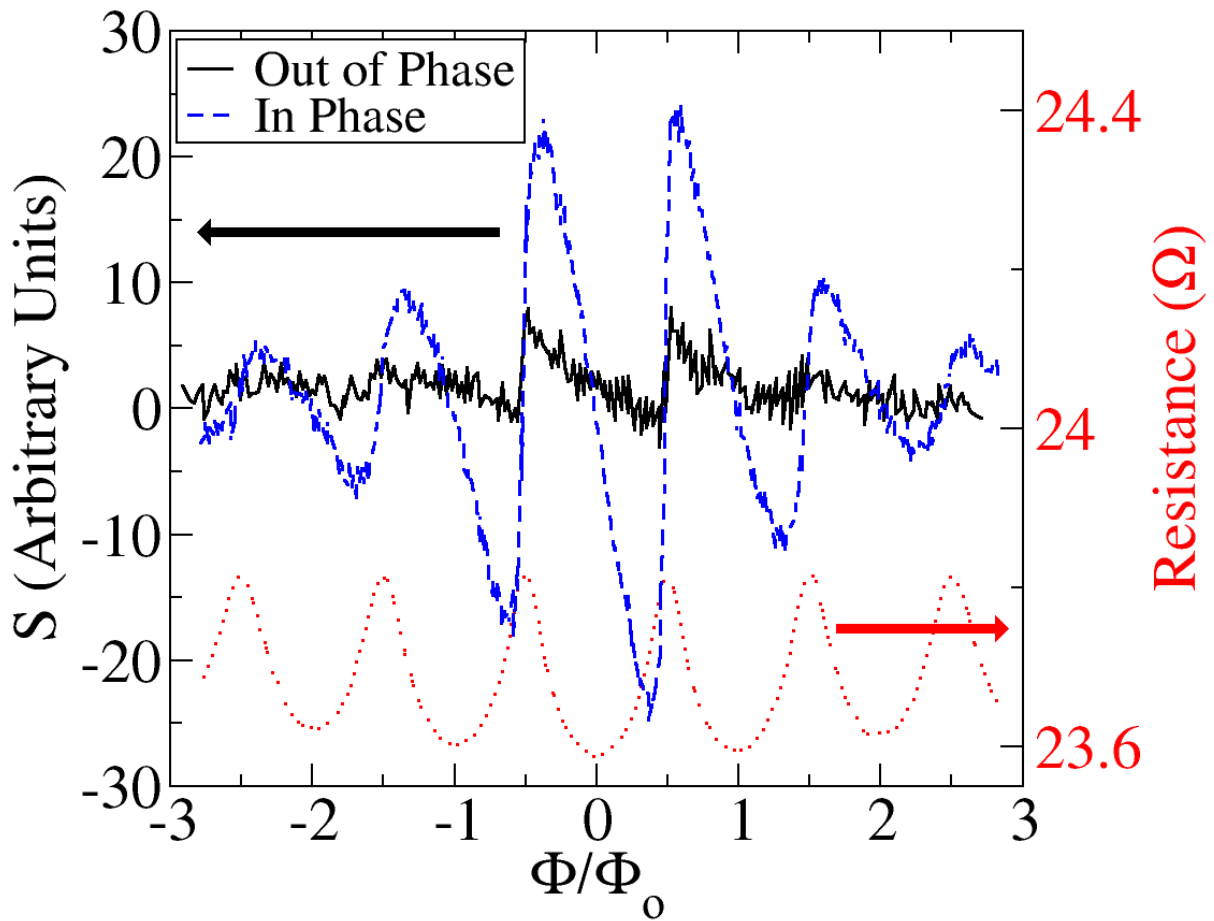}
\caption{Thermopower oscillations of the offset double-loop Andreev interferometer at T = 20 mK as a function of flux for the out-of-phase (solid line) and in-phase (dashed line) configurations of Fig. \ref{fig3}.  The flux is measured in units of the superconducting flux quantum $\Phi_o = h/2e$.  While both configurations are antisymmetric, the thermopower oscillations for the in-phase configuration are over four times larger than those of the out-of-phase configuration.  The dotted curve shows the resistance of the Andreev interferometer for the in-phase configuration.  Whether measured in this configuration, the out-of-phase configuration, or varying the flux through only one loop, the resistance oscillations are always symmetric.}
\label{fig4}
\end{center}
\end{figure}

As mentioned above, the thermopower of the interferometer is measured using identical fluxes through the two superconducting loops with two different polarities.  For the in-phase configuration, shown schematically in Fig. \ref{fig3}(c), the supercurrent circulates in the same direction (e.g. clockwise) in both loops, leading to equal and opposite Josephson currents along adjacent sections of the normal wire.  In the analogous configuration for the device measured in Ref. \cite{jiang2}, where the Josephson currents could travel along the same section, symmetric thermopower oscillations were seen.  Here, robust antisymmetric oscillations are seen (Fig. \ref{fig4}).  Two remarks should be made:   First, these antisymmetric oscillations for the offset loop, consistent with the parallelogram configuration, indicate that the symmetric oscillations of the in-line loops were likely due to the supercurrent intersecting the thermal gradient at two points on the normal wire, rather than supercurrents traveling on opposite paths along the thermal gradient.  Second, though one might expect from considerations of the single parallelogram interferometer geometry that the effect of two parallelograms with Josephson currents traveling in opposite directions along the normal wire would result in zero net change to the thermopower, this appears not to be the case.  While it is possible that the lack of cancellation is due to different fields from the two field coils or different couplings to the normal wire at the NS interfaces, these concerns can be addressed by examining the resistance oscillations of the interferometer for each superconducting loop individually.  The fact that the period of oscillations measured in field coil current is the same indicates the coils are identical, while the amplitude of the resistance oscillations due to the top loop are $\sim$ 35\% larger than those due to the bottom one, suggesting a modest asymmetry in the NS coupling of the two loops.  Though it may be the case that this asymmetry accounts for the lack thermopower cancellation, the results of the out-of-phase thermopower measurements make this explanation unlikely.

\begin{figure}
\begin{center}
\includegraphics[width=16.5cm]{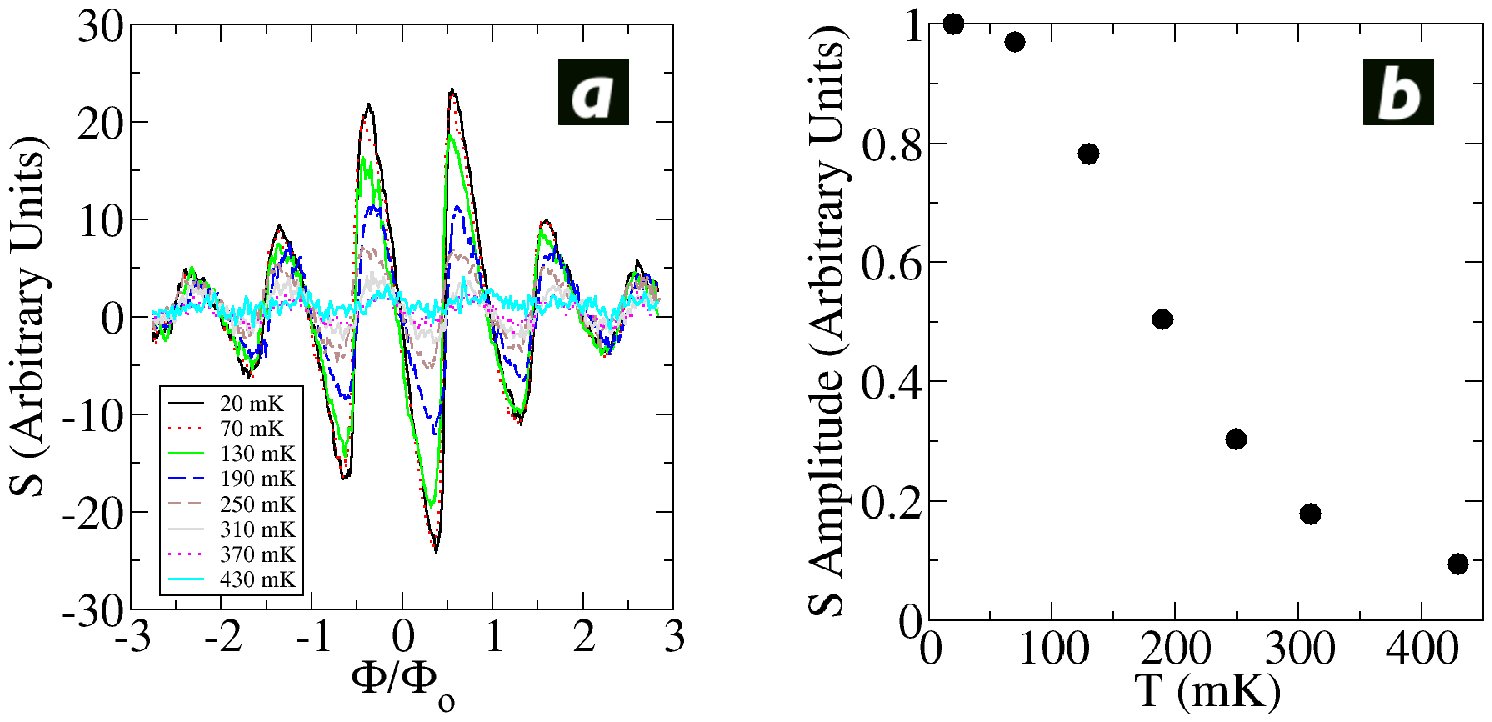}
\caption{(a) Thermopower oscillations of the Andreev interferometer as a function of flux for the in-phase configuration measured at temperatures varying from 20-430 mK.  (b)  Amplitude of these oscillations as a function of temperature.}
\label{fig5}
\end{center}
\end{figure}

The out-of-phase configuration, with Josephson currents traveling in the same direction along two different normal sections (Fig. \ref{fig3}(b)), also displays antisymmetric oscillations (Fig. \ref{fig4}), consistent with its appearance as a double parallelogram geometry.  However, the amplitude of these oscillations is over four times smaller than those measured using the in-phase configuration.  The comparative size of the two oscillations lends credence to the notion that the in-phase oscillations are not due merely to the NS interface asymmetry of the two loops.  The small relative size is possibly caused by the temperature dependence of the thermopower amplitude.  Previous experimental and theoretical work on Andreev interferometers \cite{virtanen, zou} has shown that the thermopower oscillations peak at a temperature on the order of the Thouless energy, $E_T ={\hbar{D}}/L^2$, where $L$ is the relevant length of the sample.  For $L = 1\;\mu$m, the distance between NS interfaces for a single superconducting loop, $E_c=140$ mK.  As shown in Fig. \ref{fig5} the in-phase thermopower oscillations are still reaching their peak at T = 20 mK; the out-of-phase oscillations also monotonically increase with decreasing temperature.  Since the two Josephson currents of the in-phase configuration pass along two 1 $\mu$m normal wire sections, while the Josephson current of the out-of-phase configuration passes along a 2 $\mu$m normal wire section, it may be the case that the Thouless energy and the thermopower peak in temperature are reduced by a factor of four for the out-of-phase configuration, resulting in smaller thermopower oscillations than the in-phase configuration at the same temperature.

\section{Conclusion}

We have fabricated and measured a double-loop Andreev interferometer that enables Josephson currents to travel along adjacent sections of a normal metal wire in the same or the opposite direction.  The symmetry and amplitude of thermopower oscillations observed in this device lead us to believe that symmetric thermopower oscillations observed in previously measured devices \cite{jiang2} were due to single point intersections of supercurrent and thermal current in the normal wire, and that antisymmetric thermopower oscillations result from supercurrent and thermal current traveling along the same path.  Furthermore, we have noted that the cumulative effect on the thermopower of two offset parallelogram Andreev interferometers operating in-phase does not appear to be equal to the simple cancellation of the effect of one interferometer by the other.



\begin{thebibliography}{text}

\bibitem{pothier}H. Pothier, S. Gu\'{e}ron, D. Esteve and M. H. Devoret, \textit{Phys. Rev. Lett.} \textbf{73} (1994), 2488.

\bibitem{jiang}Z. Jiang and V. Chandrasekhar \textit{Phys. Rev. B} \textbf{72} (2005), p. 020502 RC.

\bibitem{ashcroft}N. W. Ashcroft and N. D. Mermin, \textit{Solid State Physics} (New York, Holt, Rinehart and Winston, 1976).

\bibitem{eom}J. Eom, C.-J. Chien and V. Chandrasekhar \textit{Phys. Rev. Lett.}  \textbf{81} (1998), p. 437.

\bibitem{jiang2}Z. Jiang and V. Chandrasekhar \textit{Chinese J. Phys.} \textbf{43} (2005), p. 693.

\bibitem{parsons}A. Parsons, I. Sosnin and V. T. Petrashov \textit{Physica E} \textbf{18} (2003), p. 316.

\bibitem{virtanen}P. Virtanen and T. T. Heikkil\"{a} \textit{Phys. Rev. Lett.} \textbf{92} (2004), p. 177004.

\bibitem{zou}J. Zou, I. Sosnin, P. Virtanen, M. Meschke, V. T. Petrashov, and T. T. Heikkil\"{a} (2006), \textbf{cond-mat}$\backslash$0609309.

\bibitem{dikin}D.A. Dikin, S. Jung and V. Chandrasekhar \textit{Europhys. Lett.} \textbf{57} (2002), p. 564.

\end{thebibliography}
\end{document}